\documentclass[pra,twocolumn,showpacs,superscriptaddress]{revtex4}
\usepackage{graphicx}
\usepackage{amsmath}
\usepackage{amssymb}
\usepackage{epstopdf}
\usepackage{amstext}
\usepackage{latexsym}
\usepackage{pstricks}

\newcommand{\ket}[1]{\left\vert#1\right\rangle}
\newcommand{\bra}[1]{\left\langle#1\right\vert}

\begin{document}

\title{Increasing entanglement through engineered disorder in the random Ising chain}

\author{D. Binosi}
\email{binosi@ect.it}
\affiliation{European Centre for Theoretical Studies in Nuclear
  Physics and Related Areas (ECT*), Villa Tambosi, Strada delle
  Tabarelle 286, 
 I-38050 Villazzano (TN)  Italy}

\author{G. De Chiara}
\email{dechiara@science.unitn.it}
\affiliation{BEC-CNR-INFM \& Phisics Department, University of Trento,
Via Sommarive 14, I-38050 Povo (TN) Italy}
\author{S. Montangero}
\email{monta@sns.it}
\affiliation{NEST-CNR-INFM \& Scuola Normale Superiore, P.zza dei Cavalieri 7
56126 Pisa Italy
}

\author{A. Recati}
\email{recati@science.unitn.it}
\affiliation{BEC-CNR-INFM \& Phisics Department, University of Trento,
Via Sommarive 14, I-38050 Povo (TN) Italy}

\date{\today}

\begin{abstract} The ground state entanglement entropy between
block of sites in the random Ising chain is studied by means of the Von Neumann
entropy. We show that in presence of strong correlations between the
disordered couplings and local magnetic fields the entanglement
increases and becomes larger than in the ordered
case. The different behavior with respect to the uncorrelated disordered 
model is due to the drastic change of the ground state properties. 
The same result holds also for the random $3-$state quantum Potts model.
 \end{abstract}

\pacs{75.10.Pq; 03.67.Mn; 75.10.Nr}

\maketitle
\thispagestyle{empty}

Entanglement plays a central role in modern
quantum mechanics. Having been regarded for long time as the root
of the incompleteness of quantum mechanics and the source of
several paradoxes \cite{epr}, with the advent of Quantum
Information (QI) theory it has received renewed attention, and
acquired the status of a fundamental resource for QI processing
\cite{nielsen}. In the  context of strongly correlated quantum
systems entanglement arises in a natural way \cite{fazio-review},
and can in fact be regarded as a conceptual bridge
between condensed matter physics and QI theory.
In particular,  much attention has been recently devoted to the
ground state entanglement between two blocks of spins in one
dimensional spin chains
\cite{vidal02,pasquale1,salerno04,korepin,eisert05,keating05,eisler,chen04,dechiara06,weston,Skrovseth,
scudo}.  By means of both
analytical proofs and numerical simulations, this quantity has
been shown to be logarithmically divergent with the length of the
block when the spin chain is critical. Remarkably, the prefactor
of the logarithm is proportional to the central charge of the
corresponding conformal field theory (CFT) associated with the
spin model at hands \cite{hlw94}.

The average block entropy has been also studied in the context of
random critical spin
chains\cite{refael04,laflorencie05,dechiara06,santa,contin,refael07,hoyos}.
In this case, on the one hand works on random Ising and XXZ spin
1/2 chains \cite{refael04,laflorencie05,dechiara06} suggest that
the block entropy still grows logarithmically with the size of the
block but with a smaller prefactor of the logarithm with respect 
to the ordered model, being  ``renormalized''  by a factor
$\ln 2$. On the other hand,  the block entropy in random quantum 
Potts chains with spin dimension $d\ge 2$ has been studied in 
\cite{santa}, where a strong-disorder renormalization group (RG) 
analysis shows that the
average entropy still diverges logarithmically but that the prefactor of
the entropy is not proportional to the one of the corresponding
pure model and becomes larger than the latter when $d> 41$.
This provides evidence that disorder can increase entanglement with
respect to pure models, contrary to na\"{\i}ve expectations.
The same analysis have been performed in aperiodic spin
$1/2$ chains and again it has been shown that the prefactor of the
entropy logarithmic scaling is not simply the ``renormalized'' factor of the 
homogeneous model \cite{aperiodic}.
Therefore at present the possibility of associating 
an effective central charge to a non-homogeneous system is unclear. 

In this work we study a random critical Ising spin chain
\cite{fisher95} where the (nearest neighbor) couplings and the
local transverse magnetic field are drawn from the same
probability distribution and in addition share a certain (tunable)
degree of correlation. We compute the average block entropy for
this model and show that the prefactor of the entropy in a class of
random correlated chain is \emph{larger} than in the homogeneous case. 
An analogous result is found to
hold also for larger dimension as we demonstrate in the random
correlated quantum Potts model with $d=3$.

 We consider a spin chain with open boundary
conditions and with $d$ states $\vert 0\rangle, \vert 1\rangle,
\dots,\vert d-1\rangle$ per lattice site, described by the
Hamiltonian \cite{senthil} \begin{equation} {\mathcal
H}_d=-\sum_{i=1}^{L-1}J_i\sum_{n=1}^{d-1}\left(\bar S_i^z
S^z_{i+1}\right)^n-\sum_{i=1}^{L}h_i\sum_{n=1}^{d-1}\Gamma_i^n.
\label{PHam} \end{equation} Here $L$ is the length of the chain,
$\Gamma$ represents the ladder operator  $\Gamma\vert
s\rangle=\vert (s+1) \mod d\rangle$,
$\bra{k}S^z\ket{k'}=e^{2i\pi k/d}\delta_{kk'}$
($k,k'=0,\dots,d-1$) and $\bar S^z$ is the hermitian conjugate of
$S^z$.

For $d=2$, $S^z\equiv\sigma^z$ (with $\sigma^{(x,y,z)}$ the Pauli
matrices) and the model reduces to the (random) transverse field
Ising model \begin{equation} {\mathcal
H}_2=-\sum_{i=1}^{L-1}J_i\sigma_i^z
\sigma^z_{i+1}-\sum_{i=1}^{L}h_i\sigma^x_i. \end{equation} In the
pure case $J_{i}=h_i=1$ while in the random case $J_{i}$ and $h_i$
are positive random numbers drawn from a joint distribution
$\mathcal P_\alpha (J,h)$ defined in the interval $[0,1]^{\times
2}$ such that the marginal probability distributions
\begin{eqnarray}
P(J)&=&\int_0^1 dh \mathcal P_\alpha(J,h)\\
P(h)&=&\int_0^1 dJ \mathcal P_\alpha(J,h) \end{eqnarray} are both
uniform distributions in the interval $[0,1]$. This ensures that
the random models we consider are critical \cite{fisher95}. The
correlation coefficient characterizing the joint distribution
$\mathcal P_\alpha$ is defined as $\alpha=\langle(J-\langle
J\rangle)(h-\langle h\rangle)\rangle / (\sigma_J \sigma_h) $ where
$\sigma_J$ ($\sigma_h$) is the standard deviation of
$J$ ($h$). In the following we will consider the
interval $0\le\alpha\le 1$ whose endpoints correspond to
independent and perfectly correlated couplings and transverse
fields, respectively.

We study the ground state entanglement between a block of the
first $\ell$ spins and the rest of the chain. This is quantified
by the von Neumann entropy of the reduced density matrix of the
block: \begin{eqnarray}
S_\ell &=& -\textrm{Tr} \rho_\ell \log_2 \rho_\ell \\
\rho_\ell &=& \textrm{Tr}_{L-\ell} \ket{\psi_G}\bra{\psi_G}
\end{eqnarray} with $\ket{\psi_G}$ the ground state of the
Hamiltonian. In the random case we average over $N$ independent
disorder realizations.

For homogeneous models, CFT calculations show that the entropy scales
as \cite{pasquale1}:
\begin{equation} {\mathcal S}_\ell=\frac
\kappa6\log_2\left[\frac L\pi\sin\left(\frac\pi
L\ell\right)\right]+A, \label{VNentropy} \end{equation} where
$\kappa$ is equal to the central charge $c$ of the corresponding
theory and $A$ is a non-universal constant \cite{pasquale1, nonuniv}. The central charge
corresponding to the homogenous quantum Potts model depends on the local site dimension $d$ as
 \begin{equation} c_d^{\mathrm h} =
2\frac{d-1}{d+2}, 
\label{cch} \end{equation}
 \cite{santa}. 
For general non-homogeneous models, as said before, the entropy still
grows logarithmically but with a different prefactor $\kappa \ne c$ 
\cite{refael04,santa,contin}. 

The $d=2$
model can be solved analytically in both the homogeneous and
disordered case as it reduces, as noticed earlier,
to the transverse field Ising model \cite{LSM,pfeuty}:
one first performs a Jordan-Wigner
transformation that maps the spins into fermions, then  
using a Bogoliubov transformation,
whose coefficients are found via an exact diagonalization, the
system is mapped into non-interacting fermions. Finally, the
eigenvalues of the reduced density matrix $\rho_\ell$ and thus the entanglement 
can be determined from the two-point
regular and anomalous correlation functions as shown in \cite{peschel1}.

For systems with higher dimensionality one cannot resort on any
analytical tools, and in order to study the $3-$states quantum
Potts model we turn to the finite size density matrix
renormalization group (DMRG) method
\cite{white,schoelwock05,dummy}. The strategy behind the DMRG
algorithm is to construct a system block and then recursively
enlarge it, until the desired system size is reached. At every
step the basis of the corresponding Hamiltonian is truncated, so
that the size of the Hilbert space is kept manageable as the
physical system grows. The truncation of the Hilbert space is
performed by retaining the eigenstates corresponding to the $m$
highest eigenvalues of the block reduced density matrix. Several
sweeps of the finite system DMRG are performed to increase the
accuracy for non-homogeneous chains.

\begin{figure} 
\includegraphics[width=8cm]{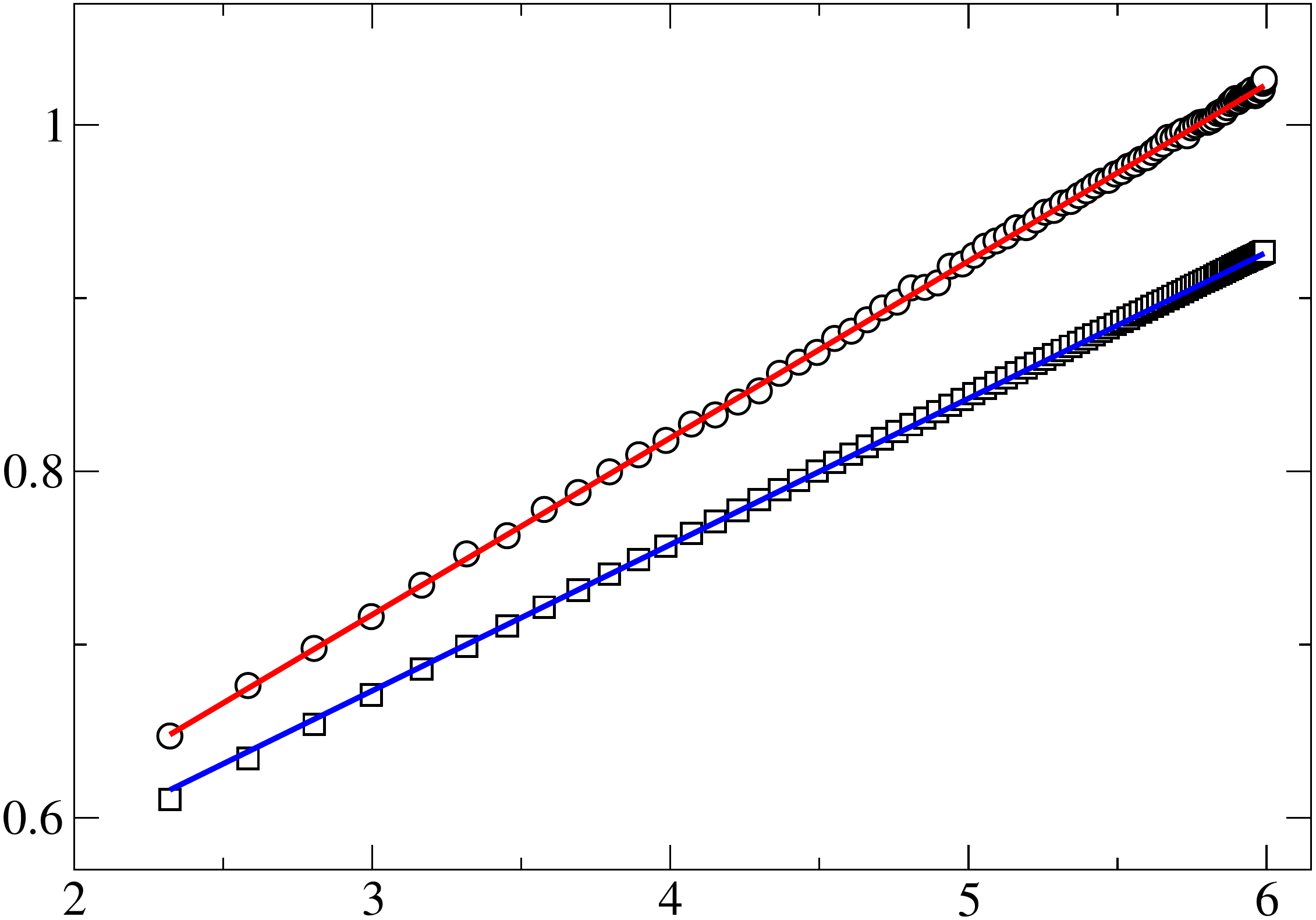}
\vspace{0.4cm}
\rput{0}(-4,-0.4){\large{$\log_2\left[\frac L\pi\sin\left(\frac\pi
L\ell\right)\right]$}} \rput{90}(-8.4,3){\large{${\mathcal S}_\ell$}}
\caption{(color online) Block entropy (in bits) averaged over
$N=5\times10^3$ configurations for an Ising spin chain of $L=200$
sites in the homogeneous (squares) and perfectly correlated random
(circles) cases. Best-fit lines with expression \eqref{VNentropy}
are also shown.} \label{entangpotts2} 
\end{figure}

{\em d=2 results: Ising model --}  In Fig.\ref{entangpotts2} we
show the results for the block entropy evaluated for the
homogenous and perfectly correlated random ($\alpha=1$,
$N=5\times10^3$ configurations) Ising chains of total length
$L=200$. Fitting the numerical data with expression
\eqref{VNentropy} we obtain for the homogeneous system
$\kappa_2^\mathrm{h}=0.50\pm0.01$, in perfect agreement
with Eq.(\ref{cch}) which gives $c_2^\mathrm{h}=0.5$ for the
homogeneous chain with $d=2$. We obtain $\kappa^{\alpha=1}_2=0.61\pm0.01$
for the perfectly correlated $\alpha=1$ case: Notice that, 
differently from previous studies on
random spin one half chains,  $\kappa^{\alpha=1}_2$ is \emph{larger}
than the prefactor for the homogeneous case resulting (in the thermodynamic limit) in an increment of the entanglement in the disordered chain with respect to the homogeneous
case \footnote{We have found that for the Ising model the constant $A$ is  
almost independent from $\alpha$, while for the $d=3$ Potts model (for the values of $L$ considered in this work) $A(\alpha=1)>A(\alpha=0)$.}.

\begin{figure}[!t]
\includegraphics[width=8cm]{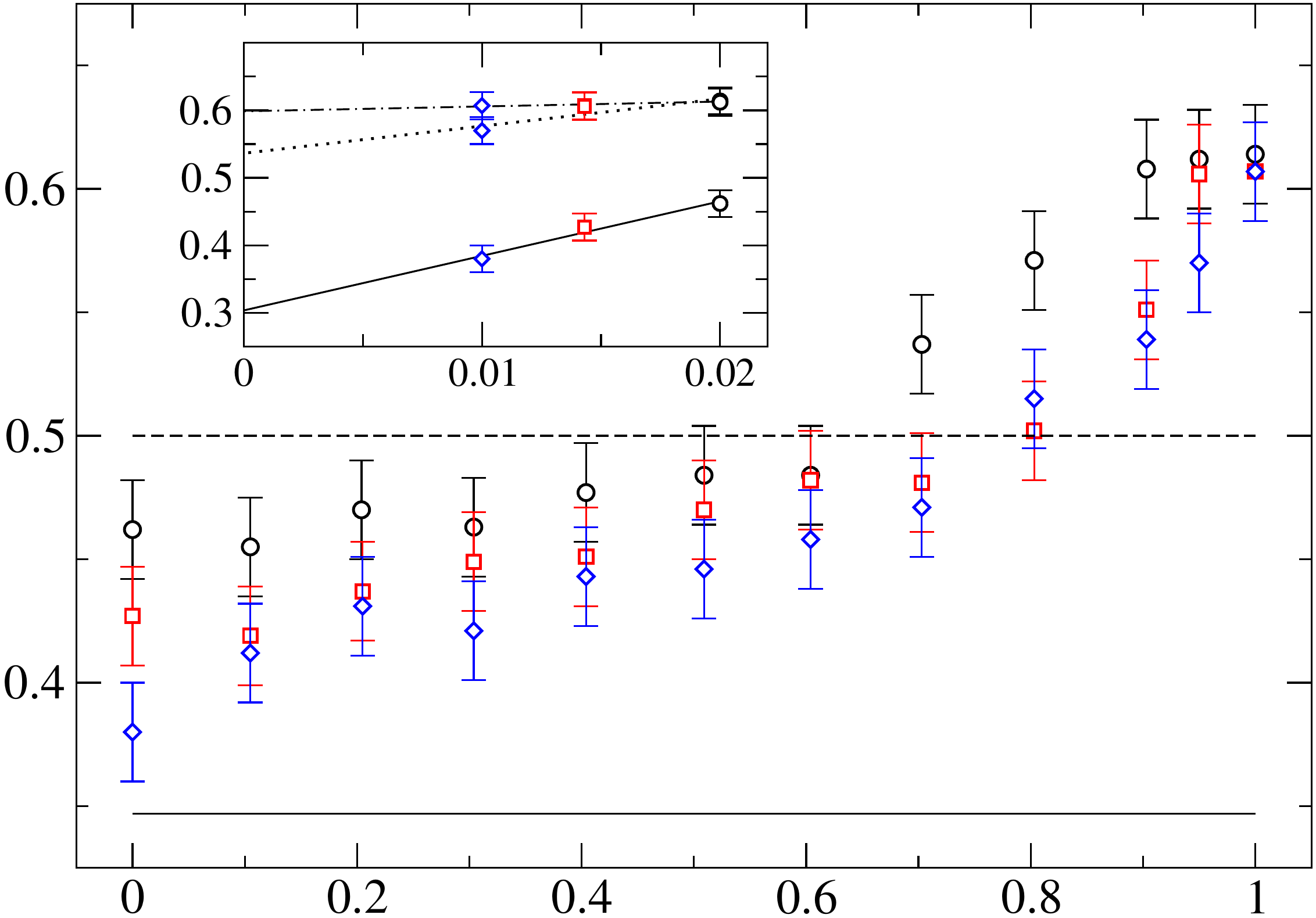}
\rput{0}(-4,-0.1){\large{$\alpha$}} 
\rput{90}(-8.4,3){\large{$\kappa_2^\alpha$}}
\rput{90}(-7.25,4.25){\normalsize{$\kappa_2^\alpha$}}
\rput{0}(-4.5,3.25){\footnotesize{$1/L$}} 
\caption{(color online) Prefactor $\kappa^\alpha_2$ of the entropy versus the
correlation coefficient $\alpha$ of the $J$ and $h$ couplings for
different chain lengths ($L=50$, black circles; $L=70$, red squares; $L=100$, blue diamonds) and $N=10^4$ configurations. The solid and dashed line gives $\kappa^{\alpha=0}_2=c_2 \ln
2\approx 0.35$ and $\kappa^h_2=c_2=0.5$ respectively.  The inset shows the
finite-size scaling of the prefactor with best-fit lines (solid, $\alpha=0$; dotted, $\alpha=0.95$; dashed-dotted, $\alpha=1$). Extrapolation to the thermodynamic limit $1/L\to0$ gives $\kappa^{\alpha=0}_2=0.30\pm0.04$ and $\kappa^{\alpha=1}_2=0.60\pm0.01$. }
 \label{cchargevscorr} \end{figure}
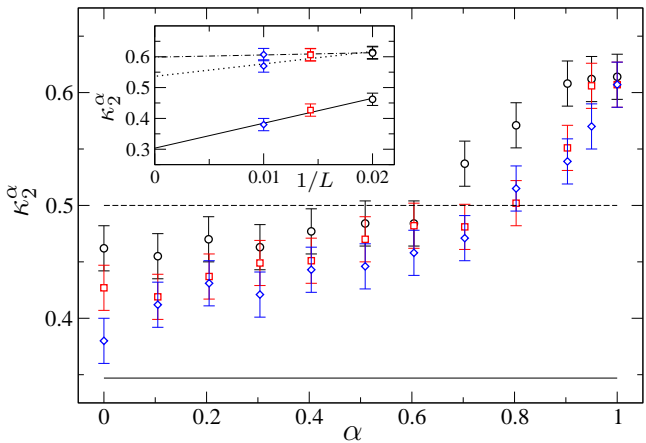

An important question that ought to be studied is the stability 
of this behavior with the correlation $\alpha$ shared by  
the couplings $J_i$ and the transverse fields $h_i$.
In Fig.\ref{cchargevscorr} we show the prefactor $\kappa_2^\alpha$
of the average entropy for different chain lengths ($L=50,\, 70,\,
100$) as a function of the correlation $\alpha$. The results show
that for small $\alpha$ we confirm the analytic result of
\cite{refael04} apart from finite size errors. Indeed, a finite size
scaling (see inset of
Fig.\ref{cchargevscorr}) gives $\kappa^{\alpha=0}_2= 0.30 \pm 0.04$ quite in 
agreement with the expected result $\kappa^{\alpha=0}_2
=c_2^{h} \ln 2\approx 0.346$; in the perfect correlated case one gets instead 
$\kappa^{\alpha=1}_2=0.60\pm0.01$ in the thermodynamic limit. 
For intermediate values of $\alpha$ our results show the departure 
from the completely correlated random Ising model and for
strong enough correlations ($\alpha> \alpha^*\sim 0.9$) 
the prefactor $\kappa_2^\alpha$ become larger than the prefactor
for the pure model $\kappa_2^{\mathrm h}=0.5$. This result is a clear
evidence that it is possible to increase the entanglement adding 
disorder to an homogenous system already for small local dimension
$d=2$. Notice that the scaling of 
$\kappa_2^\alpha$ with the system size is faster for stronger
correlations. Indeed, the perfectly correlated case, $\alpha=1$, 
seems at convergence already with $L=50$ while the uncorrelated case 
is still far for the thermodynamic limit at $L=100$. This clearly
reflects the largest space of the possible configurations of the
uncorrelated case. However, the finite size scaling suggests that in the whole
range  $\alpha^* < \alpha < 1$ we have 
$\kappa_2^\alpha > \kappa_2^h = c_2$ in the thermodinamical limit.

\begin{figure}[!t] 
\begin{tabular}{cc}
\includegraphics[width=4.2cm]{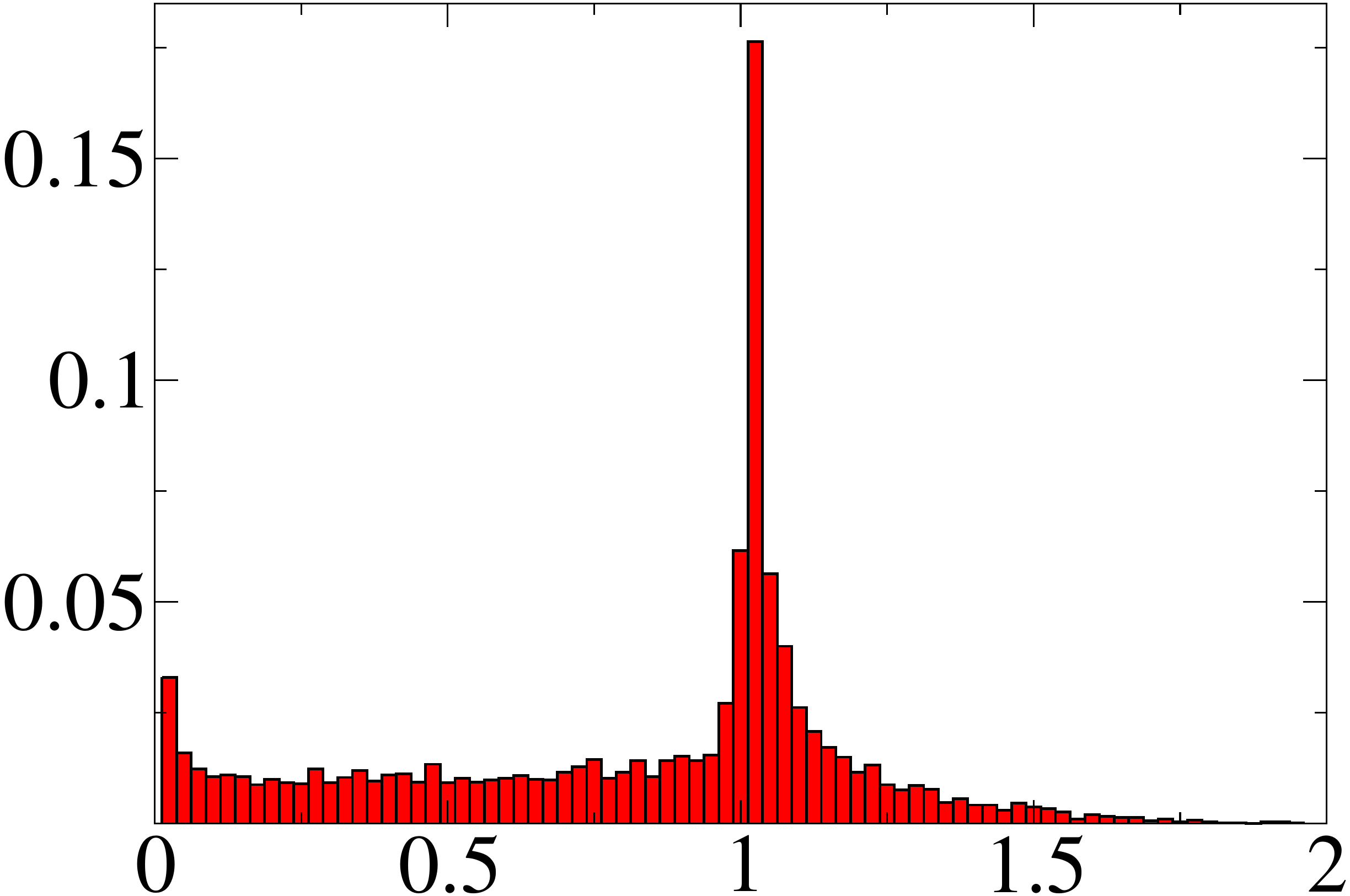}
\rput{0}(-2.1,-0.4){{\normalsize $S_{50}$}}
\rput{0}(-4.4,1.4){{\normalsize \it P}} &
\includegraphics[width=4.2cm]{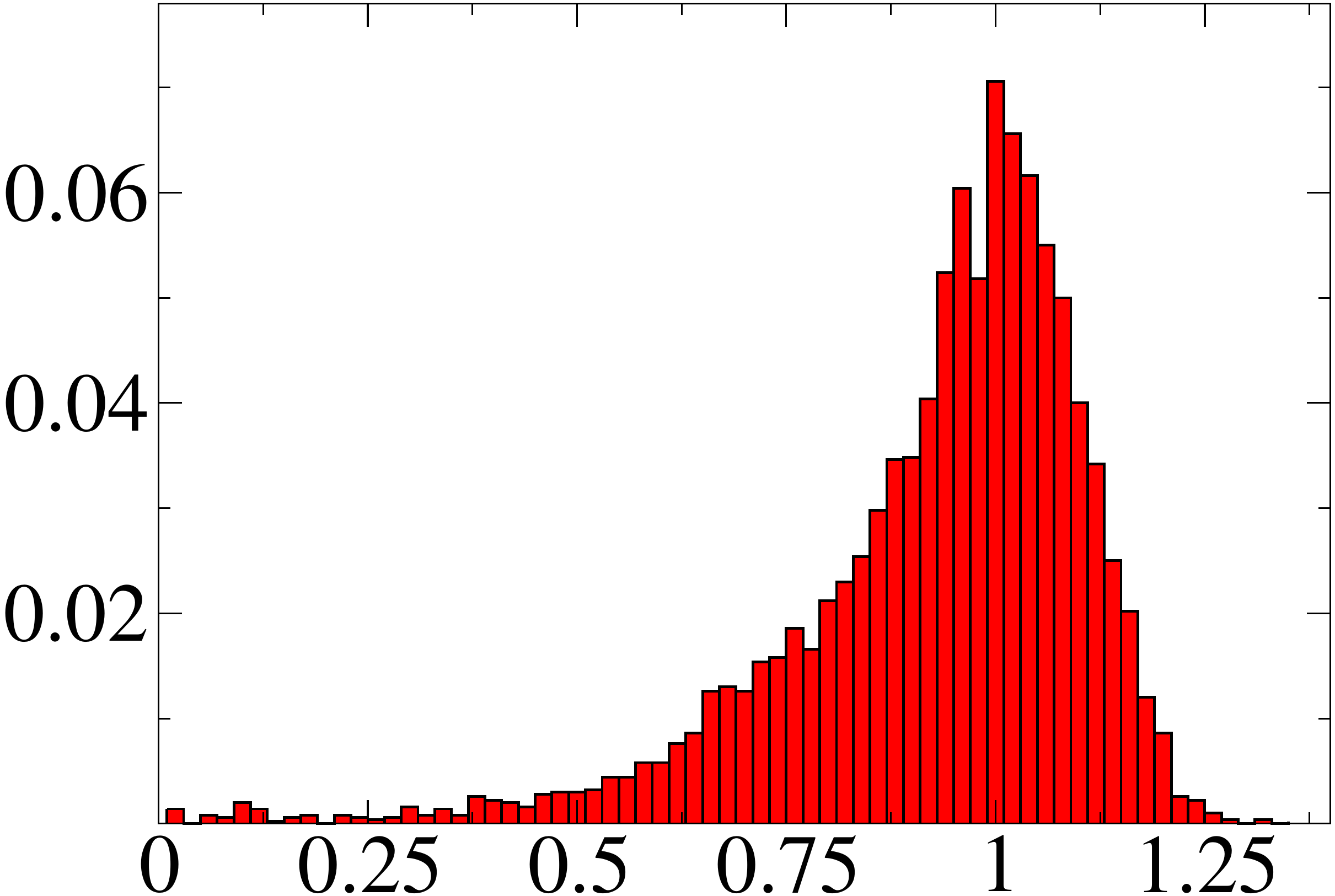}
\rput{0}(-2.1,-0.4){{\normalsize $S_{50}$}}
\rput{0}(-4.48,1.4){{\normalsize \it P}} \end{tabular}
\vspace{0.3cm}
\caption{(color online) Probability distribution for the von Neumann block
entropy for a block of length $\ell=50$ of an $L=100$ chain for
$N=5\times10^3$ configurations. Left panel: $\alpha=0$ (random
uncorrelated case); right panel: $\alpha=1$ (random perfectly
correlated case.)} \label{histograms} 
\end{figure}

A deeper insight on the ways the $\alpha > \alpha^*$ case differs form the
$\alpha=0$ one, can be gained by determining the probability
distribution $P$ of the block entropy for a fixed block length in the
two cases. This procedure will in fact unravel the presence of long distance
effective quantum correlations in the chain ground state.  The resulting
distributions for $L=100$, $\ell=50$ and $N=5\times10^3$ are shown
in Fig. \ref{histograms}. From these results it emerges that the
ground states for the two models are structurally different: for
$\alpha=0$ a single narrow peak at $S_{50}=1$ denotes the signature
of the random singlet-like phase \cite{refael04,laflorencie05}; 
for $\alpha=1$ the peak at $S_{50}=1$ is wider indicating the
occurrence of a more complex structure. This is a clear signature
that the clustering procedure described in \cite{refael04}
breaks down. Indeed, the strong disorder renormalization procedure
rely on the renormalization of the strongest bonds or local field,
obtaining  either a ferromagnetic cluster or a frozen spin. 
However, if the magnetic
field and the coupling are always of the same order
this renormalization procedure may not be applied.
It seems there exists a crossover between the two extreme regimes, namely, the random-singlet phase and the delocalized one (see Fig. \ref{histograms}).  

\begin{figure}
\includegraphics[width=8cm]{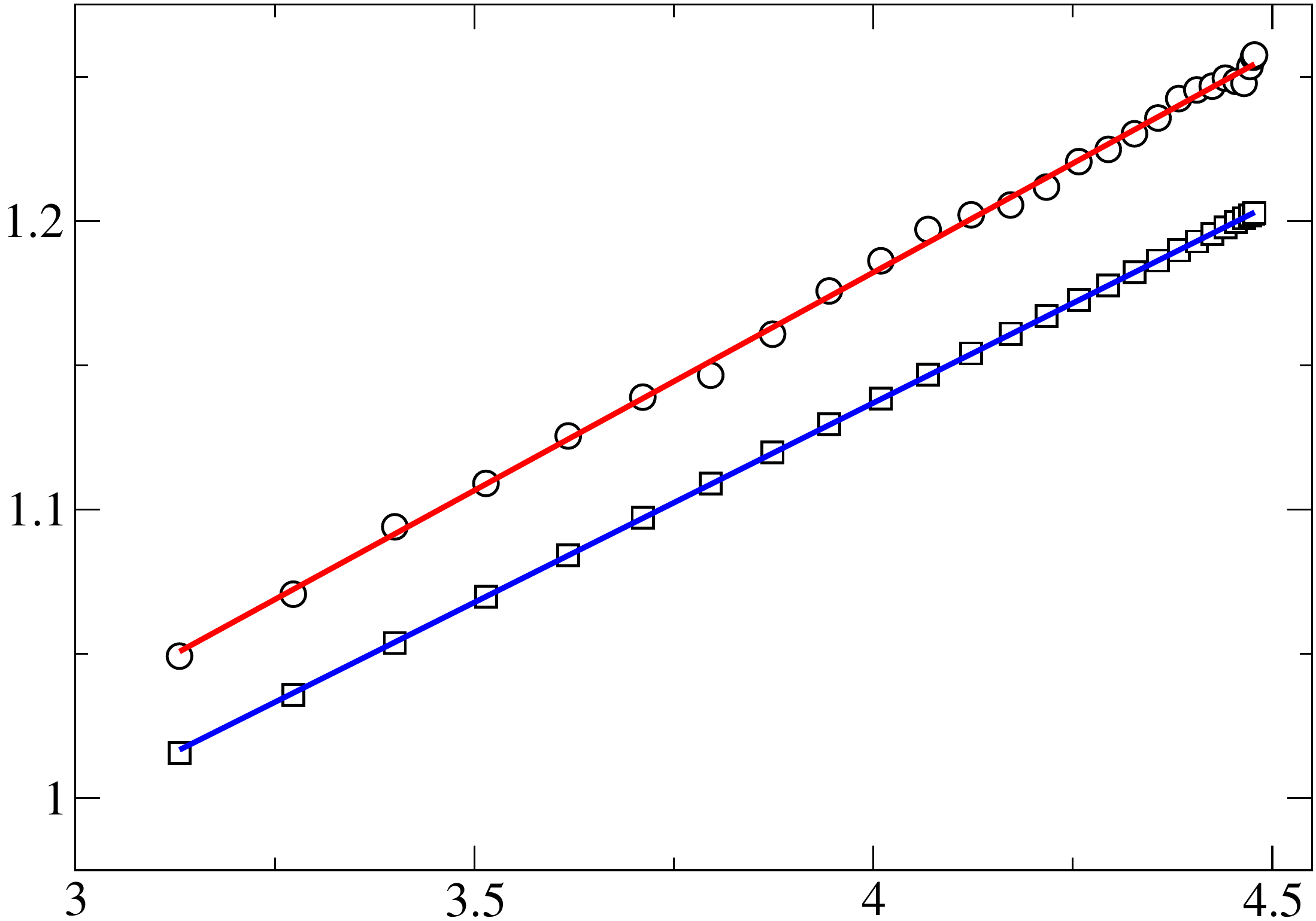}
\vspace{0.4cm}
\rput{0}(-4,-0.4){\large{$\log_2\left[\frac L\pi\sin\left(\frac\pi L\ell\right)\right]$}}
\rput{90}(-8.4,3){\large{${\mathcal S}_\ell$}}
\vspace{0.2cm}
\caption{(color online) Block entropy (in bits) averaged over $N=3\times10^3$ configurations
for a $d=3$ Potts model of $L=70$ sites in the homogeneous (squares)
and perfectly correlated random (circles) cases.  Best-fit lines with expression \eqref{VNentropy}
are also shown.} 
\label{entangpotts3}
\end{figure}

{\em d=3 results: Potts model --} 
In Fig.\ref{entangpotts3} we show the block entropy for $L=70$ in the
homogeneous and perfectly correlated random cases for $d=3$, evaluated
through a finite size DMRG algorithm using $m=30$ and
$N=3\times10^3$ configurations.  With these parameters the maximum
error in the entropy of a single configuration is less than $5\times10^{-3}$. The fit of the data for the pure model
according to expression \eqref{VNentropy} gives
$\kappa_3^\mathrm{h}=0.83\pm0.02$, in good agreement with  
value predicted by the CFT formula of  Eq.(\ref{cch}) $c_3=4/5=0.8$.
In the disordered case we find a similar picture as for the $d=2$ case:
in the perfectly correlated random case we obtain a bigger value 
of the prefactor $\kappa^{\alpha=1}_3=0.91\pm0.02 > \kappa_3^\mathrm{h}$. 
For $\alpha=0$ it is
lower than that for the homogeneous model $\kappa^{\alpha=0}_3=\ln3/2\approx0.55$
(computed in \cite{santa} via the strong disorder renormalization group).

{\it Conclusions --} In this paper we have considered a class of
random critical spin chains described by the Hamiltonian of
Eq.(\ref{PHam}). We concentrated on two models: the random Ising
model with $d=2$ and the quantum Potts model with $d=3$:
the couplings $J$ and the local transverse magnetic fields
$h$ are random but share a given amount of correlations. 
Our analysis shows that whenever
$J$ and $h$ are correlated the prefactor of the Von
Neumann entropy is always greater than the prefactor for the
uncorrelated case. Moreover, when the correlations are above a
certain threshold $\alpha^*$ the prefactor of the random correlated model is
larger than the prefactor for the homogenous chain.  It would be
interesting to confirm our numerical results applying a strong
disorder renormalization procedure.
In this work we first demonstrate how to increase
entanglement by putting static disorder in systems with small local
dimension. It is not needed to consider systems with high local
dimensions $d>41$ as in \cite{santa}:
our results indeed shows that exists a class of disordered spin one-half
models where the ground state entanglement is {\emph larger} 
than in the correspondent translational-invariant ones. 
Notice that this random correlated model could be
realized experimentally in engineered quantum systems, e.g. in 
optical lattices, ion traps or arrays of Josephson junctions.
We conclude observing that the correlated random models studied in
this work lay outside the random singlet-like phase for which a
generalized c-theorem holds. It is thus ambiguous the association
of a renormalized central charge to general non-homogeneous
systems.
 \acknowledgments We thank L. Amico, P. Calabrese,
M. Cozzini, R. Fazio, R. Santachiara for valuable discussions. GDC
acknowledges support by the European Commission through the
FET/QIPC Integrated Project  SCALA. SM acknowledges support from IST-EUROSQIP 
and the Quantum Information program of ``Centro De Giorgi'' of 
Scuola Normale Superiore. DMRG simulations have been performed using 
the code released within the IBM Faculty Awards ``Powder with Power'' 
project (\verb|http://www.qti.sns.it|). The computations have been
performed on the HPC facility of the Department of Physics,
University of Trento and on the BEN supercluster of the ECT*
centre.

\end{document}